\documentclass{PoS}
\pdfoutput=1
\title{Searches for Beyond the Standard Model Physics at the LHC: Run1 Summary and Run2 Prospects}

\ShortTitle{Searches for BSM Physics at the LHC: Run1 Summary and Run2 Prospects}

\author{\speaker{Altan Cakir}\\
       {\rm On behalf of the CMS and ATLAS Collaborations}\\
       Istanbul Technical University (ITU), Physics Enginneering Dept., 34469, Istanbul, Turkey\\
       Deutsches Elektronen-Synchrotron (DESY), 22607, Hamburg, Germany\\
       E-mail: \email{cakir@cern.ch}}

\abstract{The search for new physics is a major goal of the LHC physics program. As excitement grows for the upcoming start of Run 2, I review the CMS and ATLAS searches for physics beyond the Standard Model from Run 1 and present recent analyses. These searches have covered a wide range of new physics scenarios including Supersymmetry, new resonances, additional Higgs bosons, new hidden sectors, other Dark Matter, and multi-charged particles. In addition to reviewing some of the techniques that made the analyses possible, I will summarize what we have learned from the results and briefly discuss prospects for Run 2.}

\FullConference{Flavor Physics and CP Violation Conference\\
		 24-29 May, 2015\\
		 Nagoya, Japan}

\begin{document}
\section{Introduction}

The discovery of a Higgs boson marks a big impact of experimental and theoretical particle physics. Nevertheless, the discovery has been only the first hint in answering many unsolved questions in nature. We do not know how fundamental particles acquire their masses, what the origin of dark matter in the universe might be? The Standard Model (SM) does not adequately explain these fundamental questions. During the last two decades, advanced developments in particle physics lead to important progress of fundamental physics phenomenology that can be probed at the Large Hadron Collider (LHC).  A wide range of searches for beyond the standard model (BSM) physics has been performed at the CMS~\cite{jinstcms} and ATLAS~\cite{jinstatlas} experiments so far, unfortunately no evidence has been observed. The following analyses represent those which have become public most recently, and different searches from both collaborations have been presented in order to demonstrate the diversity of beyond the SM work. All analyses are performed in proton-proton collisions at center-of-mass energies of 8 TeV, corresponding to integrated luminosities of 19.4/fb and/or  19.7/fb.

\section{Searches for Supersymmetry}

Supersymmetry (SUSY) is one of the well motivated theoretical model for adressing most of the SM problems~\cite{Baer:2006rs}. SUSY provides elegant solutions to the unification of the gauge interactions, radiative breaking of the electroweak symmetry and dark matter in the universe. Under the conservation of R-parity, the lightest SUSY particle (LSP) is stable and is an excellent candidate for the dark matter. Therefore, the discovery (or exclusion) of weak-scale SUSY is one of the highest physics priorities for the current and future LHC programs\footnote{https://twiki.cern.ch/twiki/bin/view/CMSPublic/PhysicsResultsSUS}$^{,}$\footnote{https://twiki.cern.ch/twiki/bin/view/AtlasPublic/SupersymmetryPublicResults}.  

\subsection{Searches for third-generation squark production in fully hadronic final states in proton-proton collisions at $\sqrt{s}$ = 8 TeV}
 
In this analysis three different exclusive searches, a multijet search requires one fully reconstructed top quark, a dijet search requiring one or two jets originating from b-tagged quarks, and a monojet channel, have been performed~\cite{hadronicsquark}. Each exclusive channel is optimized for a different decay topology corresponding the direct production of either a pair of top squarks ($\tilde{t}$$\tilde{t}$) or bottom squarks ($\tilde{b}$$\tilde{b}$)  decaying top fully hadronic final states with large transverse momentum imbalance (MET). Within the context of natural SUSY, several scenarios are taken into account. They are based on the pair production of top or bottom squarks followed by the decay of the top or bottom squarks according to $\tilde{t}$$\rightarrow$$t$$\chi^{0}_1$, $\tilde{t}$$\rightarrow$$b$$\chi^{\pm}_1$ with $\chi^{\pm}_1$$\rightarrow$$b$$W^{\pm}$,  $\tilde{t}$$\rightarrow$$b$$\chi^{0}_1$,  $\tilde{b}$$\rightarrow$$b$$\chi^{0}_1$. 

The first channel, referred as multijet t-tagged search, for pairs of hadronically decaying top quarks with large MET in the final state is examined by the scenario of top-squark pair production, assuming that the mass difference between $\tilde{t}$ and $\chi^{0}_1$ (the stable lightest supersymmetric particle - called LSP) is larger than the mass of top quark, $m_{\tilde{t}}$ - $m_{\chi^{0}_1}$ > $m_t$. The next search, which is sensitive to scenarios with large or intermediate mass differences between the bottom squark and the LSP, is the dijet b-tagged channel. This analysis requires large MET and one or two jets identified as originating from bottom quarks. The third light flavor hard jet, which may come from initial state radiation (ISR), is incorporated. The last search channel, the monojet analysis, has been used for bottom-squark pair production in compressed spectrum scenarios, which the parent sparticles are close in mass to the daughter sparticles. Small mass splitting $\Delta m$ = $m_{\tilde{t}} - m_{\tilde{\chi}^{0}_{1}}$ or $\Delta m$ = $m_{\tilde{b}} - m_{\tilde{\chi}^{0}_{1}}$ leave little visible energy in the detector, making the signal events difficult to distinguish from the SM background. However, events with an energetic ISR jet recoiling against MET from the LSP can provide a clear signal for compressed events. Therefore, a search for events with a single jet and significant MET, called as monojet channel, has been performed.

\begin{figure}[htbp]
\centering
\includegraphics[height=4.7cm]{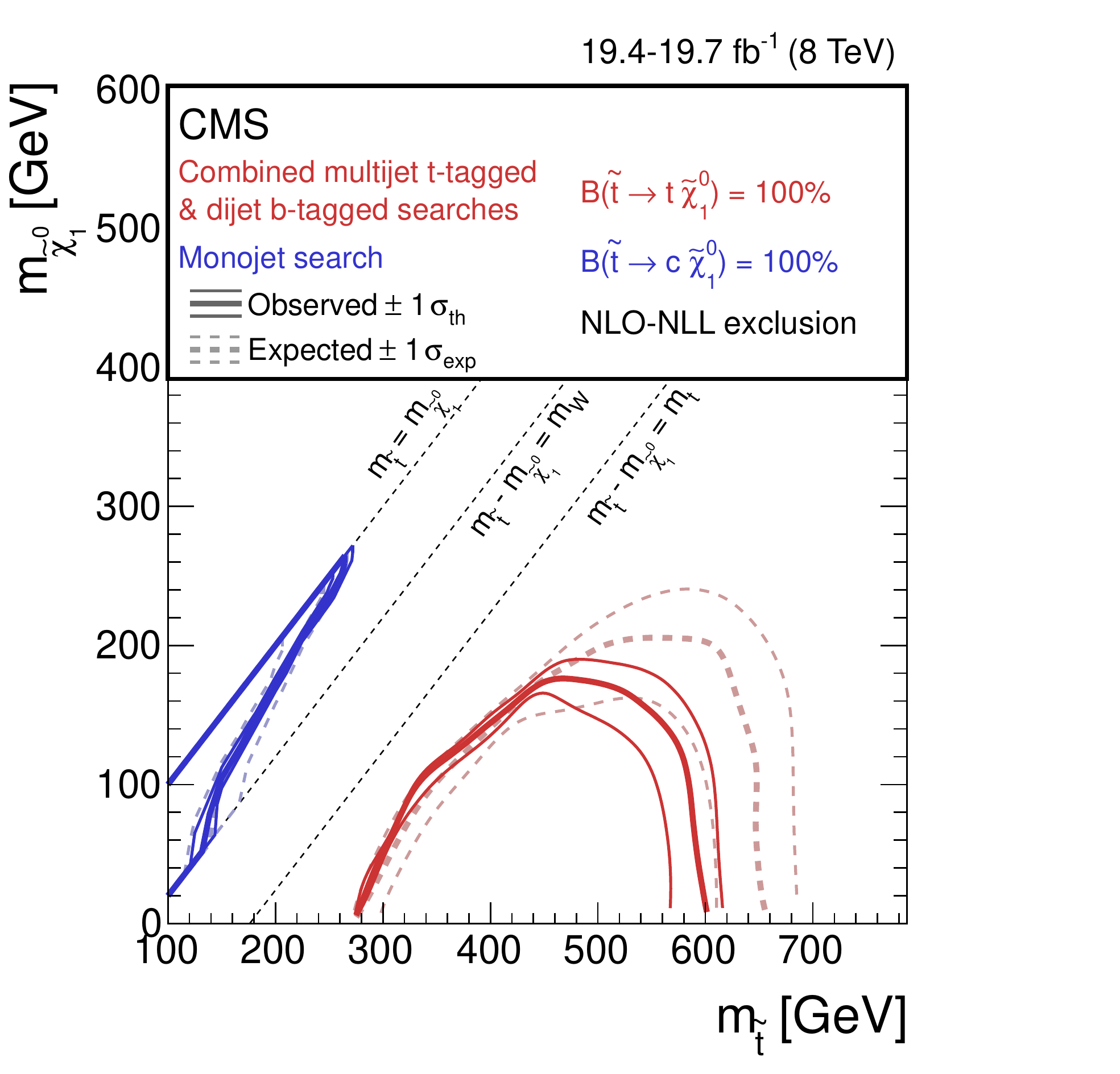}
\includegraphics[height=4.7cm]{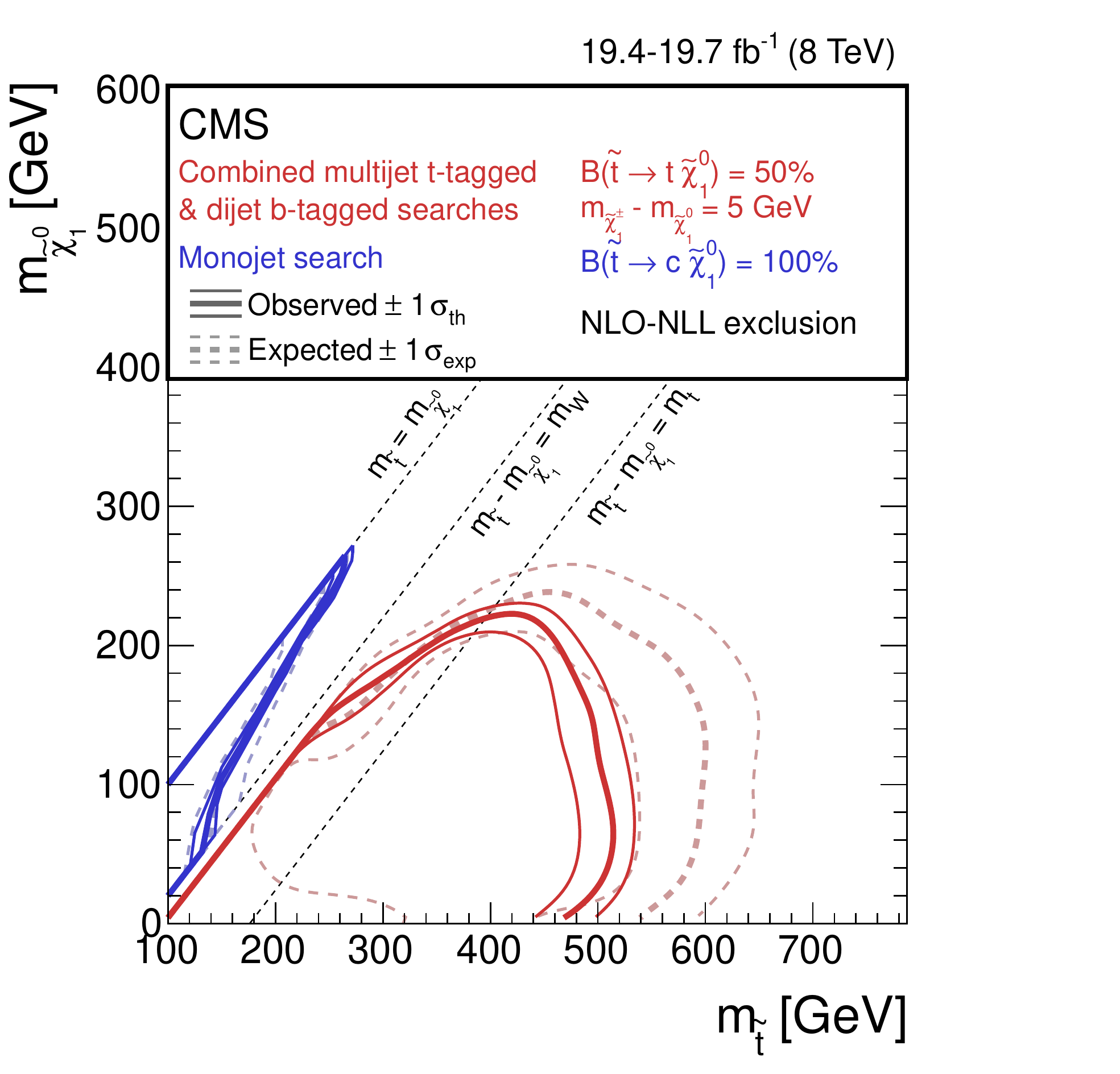}
\includegraphics[height=4.7cm]{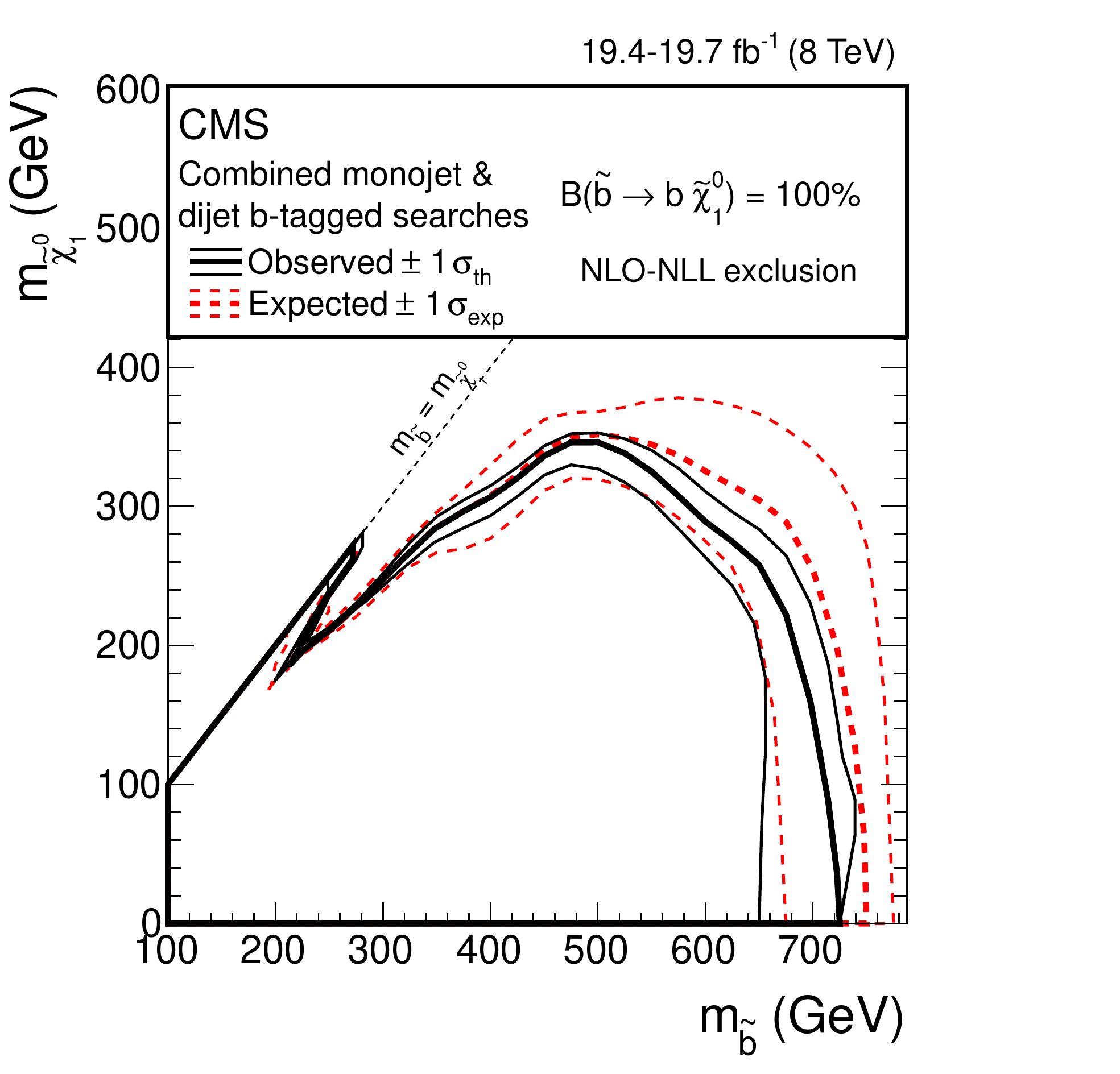}
\caption{Expected and observed 95$\%$ CL exclusion limits in the ($m_{\tilde{t}}$,  $\tilde{\chi}^{0}_{1}$) mass plane for top-squark pair production (upper-row and lower-left) assuming various branching fractions and mass splittings to the decay to $\tilde{t}$ $\rightarrow$ t$\tilde{\chi}^{0}_{1}$ (upper left), $\tilde{t}$ $\rightarrow$ b$\tilde{\chi}^{0}_{1}$ and $\tilde{t}$ $\rightarrow$ c$\tilde{\chi}^{0}_{1}$. Expected and observed 95$\%$ exclusion limits in the ($m_{\tilde{b}}$,  $\tilde{\chi}^{0}_{1}$) mass plane for bottom-squark pair production$\tilde{b}$ $\rightarrow$ b$\tilde{\chi}^{0}_{1}$ (lower-right) assuming 100$\%$brancing fraction decay to  $\tilde{b}$ $\rightarrow$ b$\tilde{\chi}^{0}_{1}$.}
\label{fig:first}
\end{figure}

Figure~\ref{fig:first} shows the 95$\%$ CL exclusion limits for top squark or bottom squark and LSP ($\tilde{\chi}^{0}_{1}$) masses, for either $\tilde{t}$ $\tilde{\bar{t}}$ $\rightarrow$ $t \bar{t}$ $\tilde{\chi}^{0}_{1}$$\tilde{\chi}^{0}_{1}$ or $\tilde{t}$ $\tilde{\bar{t}}$ $\rightarrow$ $\tilde{c}$ $\tilde{\bar{c}}$ $\tilde{\chi}^{0}_{1}$$\tilde{\chi}^{0}_{1}$, and $\tilde{b}$ $\tilde{\bar{b}}$ $\rightarrow$ $b \bar{b}$ $\tilde{\chi}^{0}_{1}$$\tilde{\chi}^{0}_{1}$, which are kinematically allowed. The black diagonal dashed lines show the various kinematic regimes for top squark  decay and bottom-squark decay. No significant excesses above the predictions from the standard model are observed, and 95$\%$ CL exclusion limits are placed on top and bottom squark masses.

\subsection{Search for supersymmetry in events containing a same-flavour opposite-sign dilepton pair, jets, and large missing transverse momentum at $\sqrt{s}$ = $8$ TeV}

In this analysis, events with leptons (e or $\mu$) in same-flavour opposite-sign (SFOS) pairs via squarks and gluino production are considered. The following SFOS processes: the leptonic decay of a Z boson, Z $\rightarrow$ $l^{+} l^{-}$, and the decay $\tilde{\chi}^{0}_{2}$ $\rightarrow$ $l^{+} l^{-}$$\tilde{\chi}^{0}_{1}$, which includes additional contributions from $\tilde{\chi}^{0}_{2}$ $\rightarrow$ $l^{\pm (*)} l^{\mp}$ $\rightarrow$ $l^{+} l^{-}$$\tilde{\chi}^{0}_{1}$ and $\tilde{\chi}^{0}_{2}$ $\rightarrow$ $Z^{*}$$\tilde{\chi}^{0}_{1}$ $\rightarrow$ $l^{+} l^{-}$$\tilde{\chi}^{0}_{1}$ are presented. In generalized gauge mediated supersymmetry breaking scenario (GGM), gravitino ($\tilde{G}$), which is defined as LSP particle, and Z boson may be produced via the following decay: $\tilde{\chi}^{0}_{1}$ $\rightarrow$ Z$\tilde{G}$. These complementary SFOS processes, in particular, are the main motivation of this analysis~\cite{Aad:2015wqa}. 

SFOS lepton production channels are examined by their distributions of dilepton invariant mass $(m_{ll})$. Z $\rightarrow$ $l^{+} l^{-}$ leads to peak in the $m_{ll}$ distribution around the Z boson mass, while the decay $\tilde{\chi}^{0}_{2}$ $\rightarrow$ $l^{+} l^{-}$$\tilde{\chi}^{0}_{1}$ leads to a rising distribution in $(m_{ll})$ that shows edge like behavior at a kinematics endpoint, since events with larger $(m_{ll})$ values would violate energy conservation in the decay of the $\tilde{\chi}^{0}_{2}$ particle.  A search for events with a SFOS lepton pair consistent with originating from the decay of a Z boson (called as on-Z search) targets SUSY models with Z boson production. A search for events with a SFOS lepton pair inconsistent with Z boson decay (called as off-Z search) targets the decay $\tilde{\chi}^{0}_{2}$ $\rightarrow$ $l^{+} l^{-}$$\tilde{\chi}^{0}_{1}$. 

\begin{figure}[htbp]
\centering
\includegraphics[height=4.5cm]{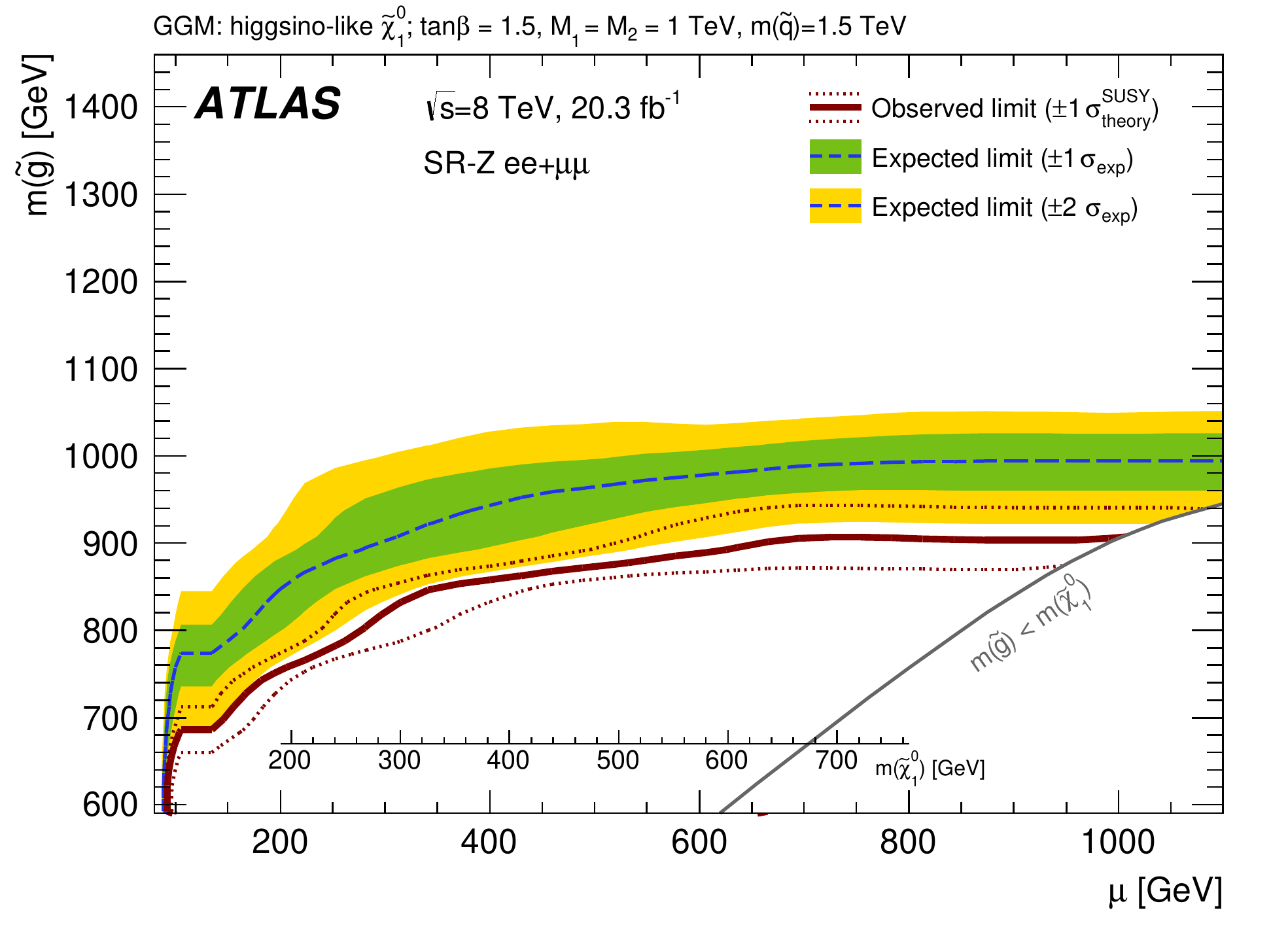}
\includegraphics[height=4.5cm]{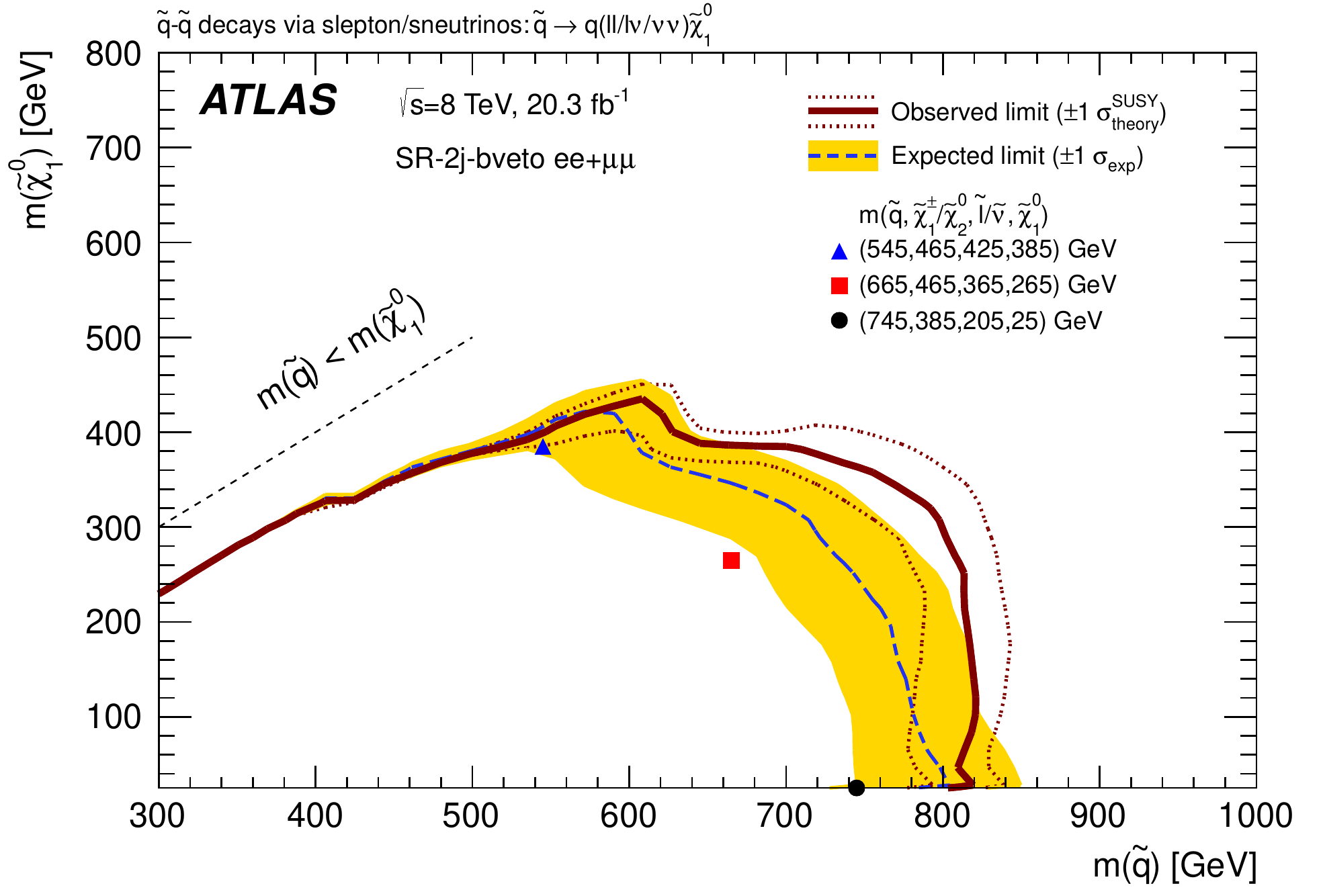}
\caption{Expected and observed 95$\%$ CL limit from the on-Z combined same-flavor channels in the $\mu$ (higgsino mass parameter) versus m($\tilde{g}$) plane in the GGM model with tan $\beta$ = 1.5 (left). Expected and observed 95$\%$ CL limit from the off-Z channels in the (top) squark - LSP mass plane (right).}
\label{fig:second}
\end{figure}


For the on-Z analysis, the data exceeds the background expectations in the ee ($\mu\mu$) channel with a significance of 3.0 (1.7) standard deviations. Exclusion limits in specific models~\cite{Aad:2015wqa} allow us to illustrate which regions of the model parameter space are affected by the observed excess, by comparing the expected and observed limits. Given the observed excess of events with respect to the SM prediction, the observed limits are weaker than expected. 

\section{Searches for Exotica}

There are many new physics models for the BSM, conventionally separated into supersymmetry models, and all other BSM models, referred to as exotica.  This provides a large terrain to probe experimentally, and both ATLAS and CMS experiments at the LHC have run extensive physics programs to cover as many scenarios as possible\footnote{https://twiki.cern.ch/twiki/bin/view/CMSPublic/PhysicsResultsEXO}$^{,}$\footnote{https://twiki.cern.ch/twiki/bin/view/AtlasPublic/ExoticsPublicResults}$^{,}$\footnote{https://twiki.cern.ch/twiki/bin/view/CMSPublic/PhysicsResultsB2G}. 

\subsection{Search for third-generation scalar leptoquarks in the t$\tau$ channel at $\sqrt{s}$ = $8$ TeV}

Leptoquarks ($LQ$), which are hypothetical particles that carry both lepton (L) and baryon (B) quantum numbers, are mostly investigated in theories beyond the SM, such as grand unification and compositeness~\cite{Chakdar:1}. In this analysis, third-generation scalar Leptoquarks ($LQ_{3}$) decaying into a top quark and a $\tau$ lepton have been presented for the first time~\cite{Khachatryan:XXX}. Previous searches at the LHC have targeted $LQ$ decaying into quarks and leptons of the first and second generations or the third-generation in the $LQ_3$ $\rightarrow$ b$\nu$ and $LQ_3$ $\rightarrow$ b$\tau$ decay channels. Furthermore, this search can also be interpreted in the context of R-parity violating (RPV) supersymmetric models where the supersymmetric partner of the bottom quark (bottom squark) decays into a top quark and a $\tau$ lepton via the RPV coupling~\cite{Barbier:2004ez}.

\begin{figure}[htbp]
\centering
\includegraphics[height=4.7cm]{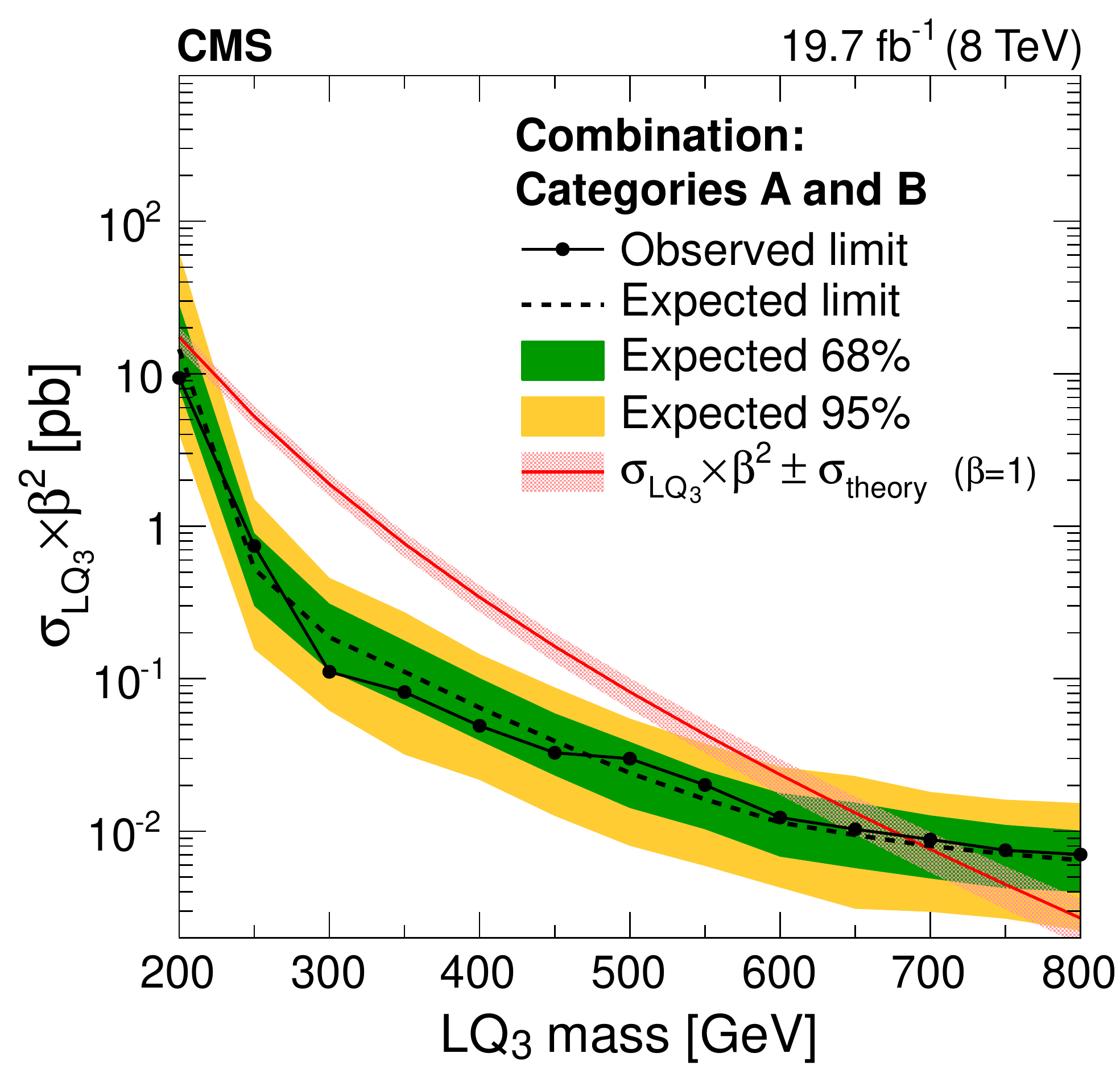}
\includegraphics[height=4.7cm]{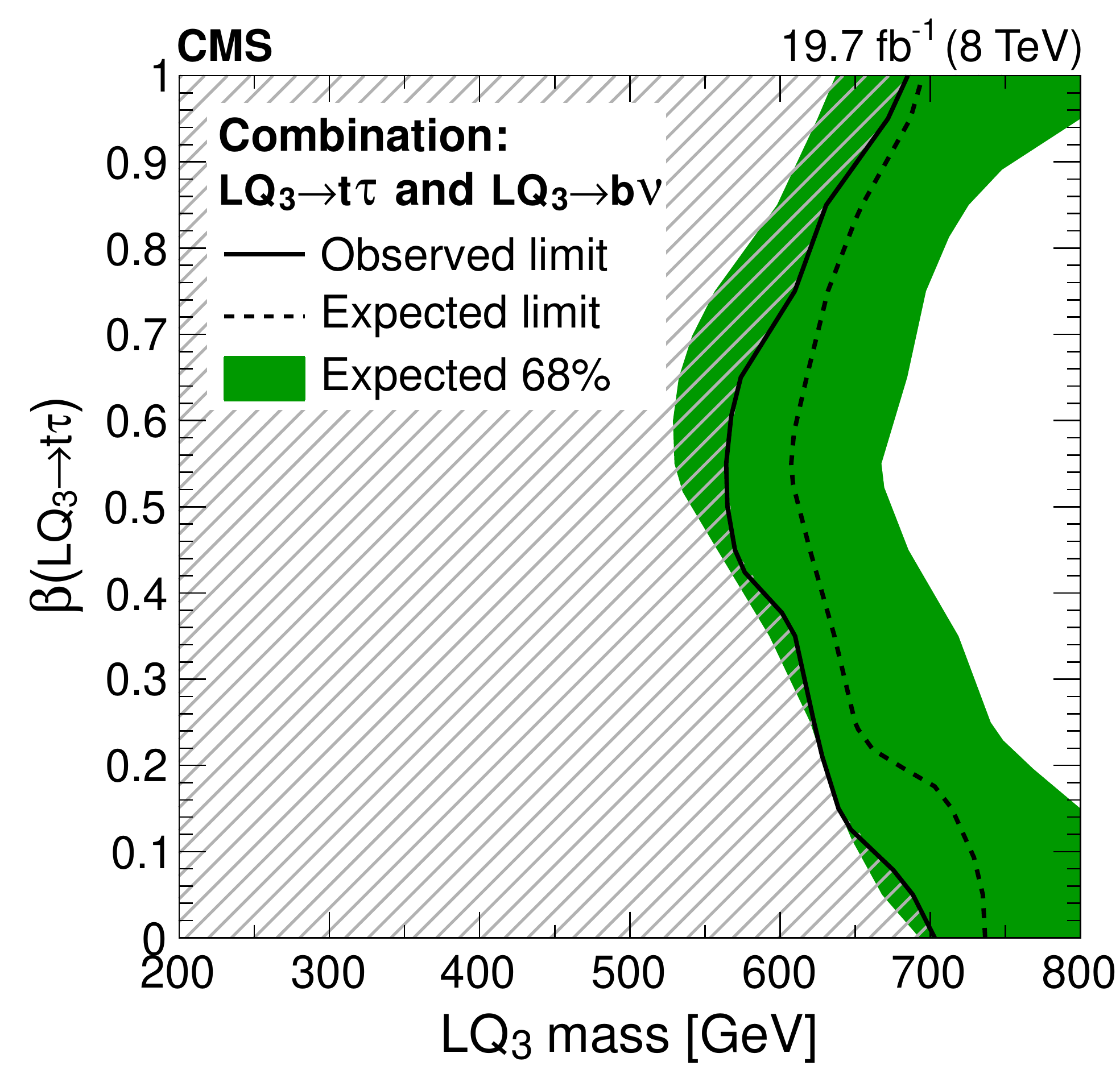}
\caption{The expected and observed exclusion limits at 95$\%$ CL on the $LQ_3$ pair production cross section times $\beta^{2}$ in two different categories (combination) (left).  The expected and observed limits on the LQ branching fraction $\beta$ as a function of the LQ mass (right). The shaded region is obtained by including the results in Ref. \cite{Khachatryan:2015wza}, reinterpreted for the $LQ_3$ $\rightarrow$ b$\nu$ scenario. }
\label{fig:third}
\end{figure}

A short summary of the search strategies, defined as event categories A and B, has given in Ref~\cite{Khachatryan:XXX}. In these categories, different MET and jet multiplicity requirements have been assigned for the muon and $\tau$ lepton candidates, which differ only in the thresholds of the isolation requirements and selection of the lepton pairs. As a result, the observed and expected exclusion limits as a function of the LQ mass are presented in Fig.~\ref{fig:third}. Assuming 100$\%$ branching fraction of LQ decays to top quark and $\tau$ lepton pairs, pair production of third-generation LQs is excluded for masses up to 685 GeV with an expected limit of 695 GeV. For bottom squarks search in RPV SUSY, third-generation scalar LQs are also excluded for masses below 700 GeV for $\beta$ = 0 and for masses below 560 GeV over the full $\beta$ range.

\subsection{Analysis of events with b-jets and a pair of leptons of the same charge at $\sqrt{s}$ = $8$ TeV}

This analysis searches for a BSM channel resulting in pairs of isolated leptons with the same electric charge, MET, and b-jets. This signal topology provides a promising channel due to the small SM yields of such events, and several types of BSM models, such as the vector-like quarks (VLQ), the four-top-quark production cross section, the existence of a fourth generation of chiral quarks, and production of two positively charged top quarks, may contribute. Various analyses techniques and background estimation methods have been extensively used in this search and further details can be followed in Ref~\cite{Aad:2015gdg}. 

\begin{figure}[htbp]
\centering
\includegraphics[height=4.7cm]{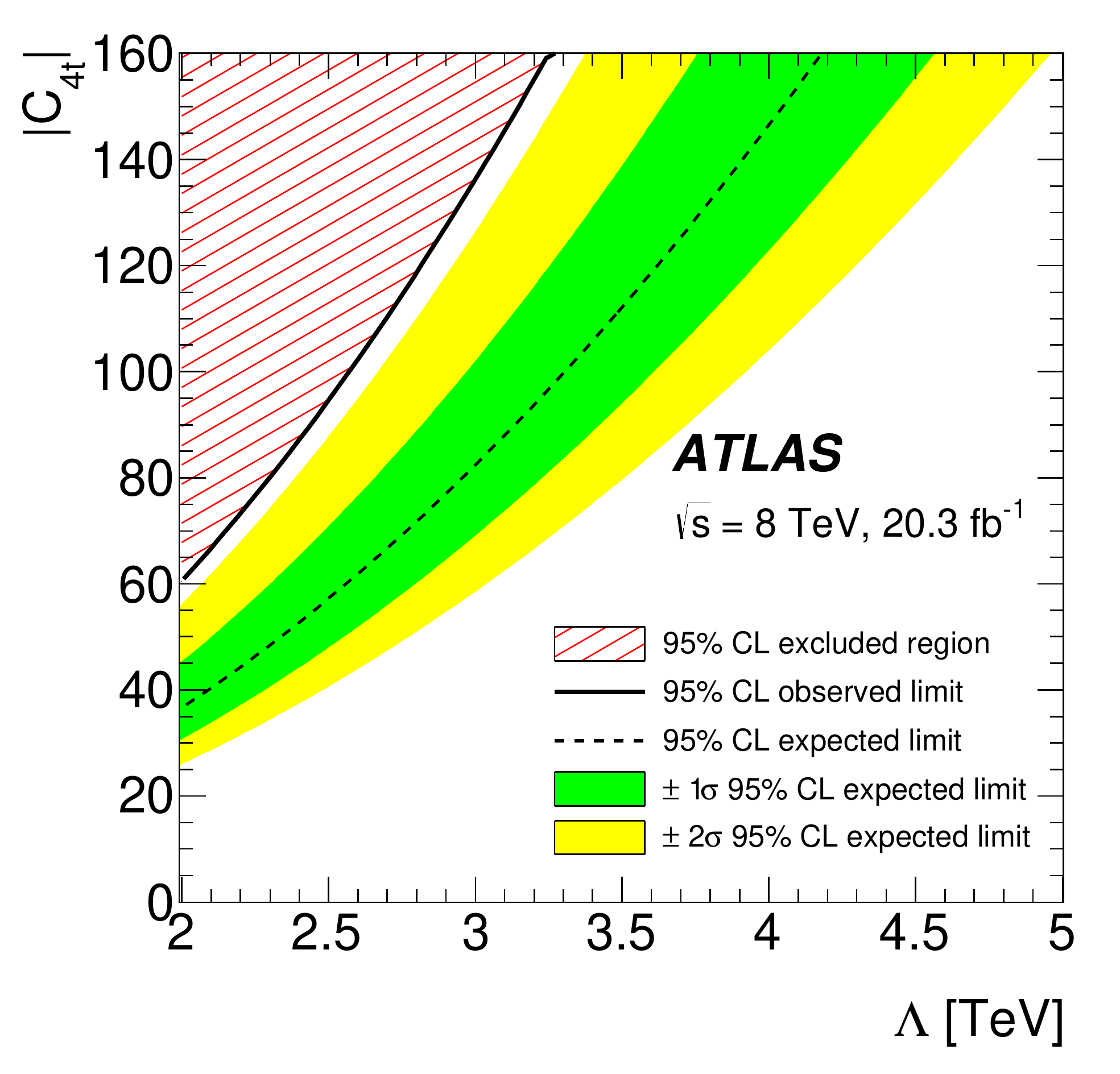}
\includegraphics[height=4.7cm]{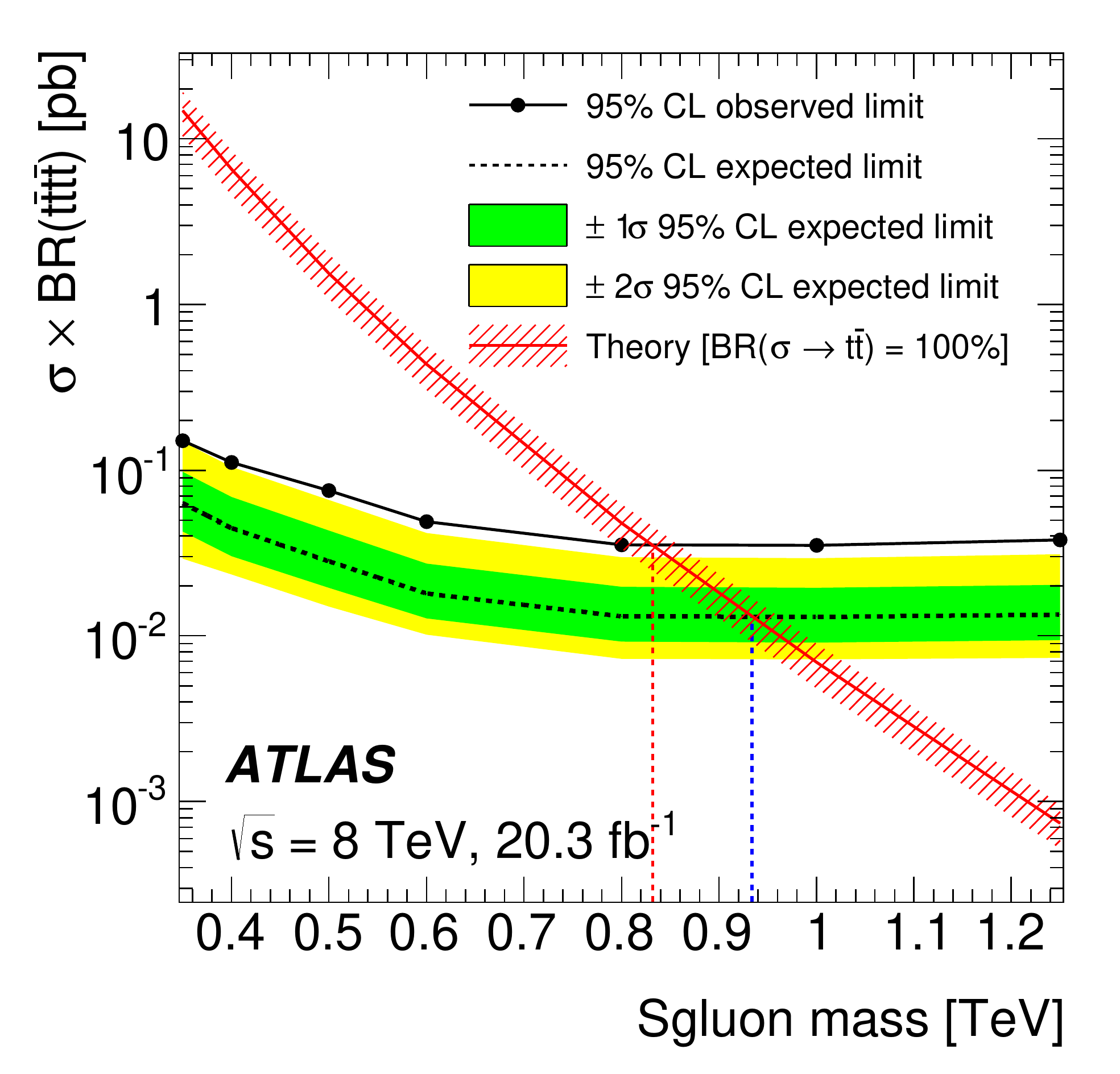}
\includegraphics[height=4.7cm]{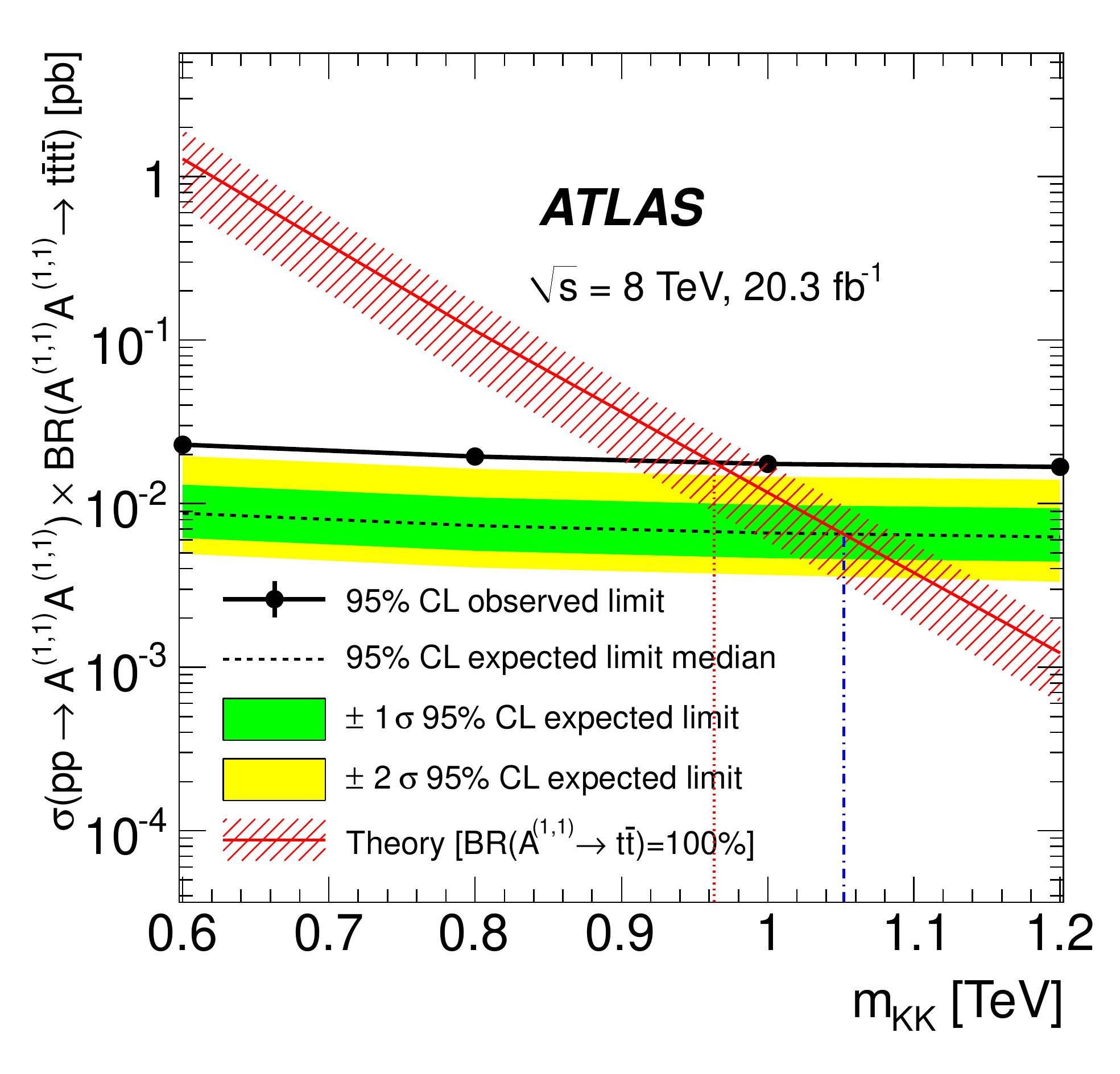}
\caption{Expected and observed limits on the coupling constant $|C_{4t}|$ in the contact interaction model for four-top production as a function of the BSM physics energy scale $\Lambda$ (left), the sgluon pair-production cross section times branching ratio to four top quarks as a function of the sgluon mass (middle), the four-top-quark production rate for the 2UED/RPP model in the symmetric case (right) with 100$\%$ branching fractions for A, which refers Kaluza-Klein (KK) excitations of the photon in theory. The vertical dashed lines indicate the expected and observed limits on the sgluon mass or on $m_{KK}$, and the shaded band around the theory cross section indicates the total uncertainty on the calculation.}
\label{fig:fourth}
\end{figure}

In general, the observed yields agree well with the expectation from background in the signal regions defined for searching for positively charged top quark pair production. In contrast, some of the signal regions, defined for VLQ, $b^{\prime}$-quark, and four-top-quark production, exhibit an excess over expected background in Fig.~\ref{fig:fourth}. The excess is largest in the subset of the signal regions used for the four-top-quark search, where at least two b-tagged jets are selected. The excess is not significant enough to support a claim of BSM physics. Therefore 95$\%$ CL upper limits on cross sections (or lower limits on masses) relevant for each model has been presented~\cite{Aad:2015gdg}. 

\subsection{Search for narrow high-mass resonances at $\sqrt{s}$ = 8 TeV decaying to Z and Higgs bosons}
This analysis focused on BSM models that predict an additional neutral gauge boson (generically called a $Z^{\prime}$) and those that incorporate composite Higgs bosons at the final state. In this search, a resonance with a mass in the range 0.8 - 2.5 TeV decaying to ZH, where the Z boson decays to qq and the Higgs boson, has been performed~\cite{Khachatryan:2015ywa}. Moreover, the directions of the particles stemming from Z and H boson decays are separated by a small angle. This feature is called as the boosted regime. For the case of Z $\rightarrow$ qq, this results in the presence of one single reconstructed jet after hadronization called a Z-jet. Thus, the novel feature of this search is the reconstruction and selection of a $\tau$ pair in the boosted regime.

\begin{figure}[htbp]
\centering
\includegraphics[height=4.7cm]{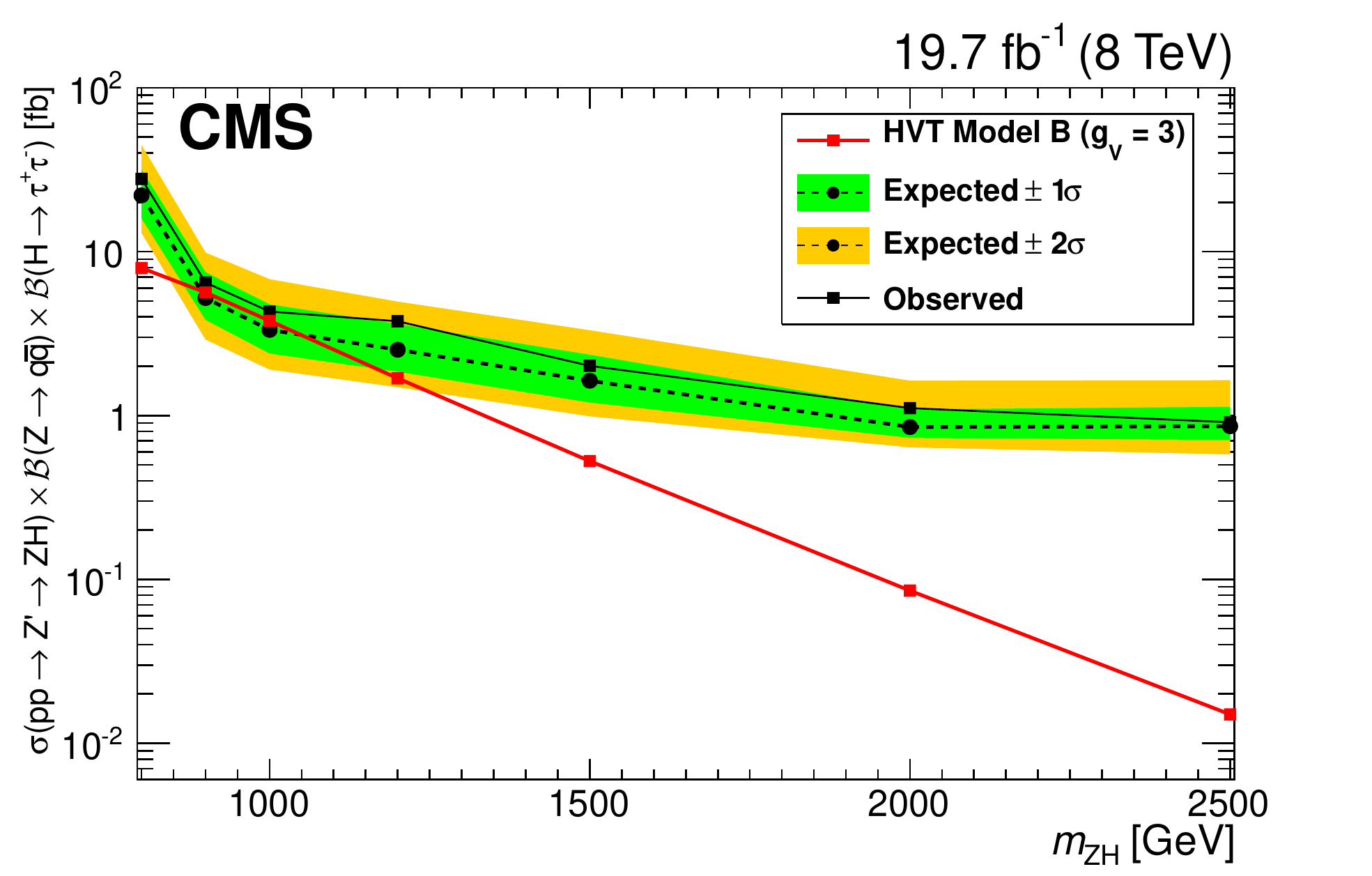}
\includegraphics[height=4.7cm]{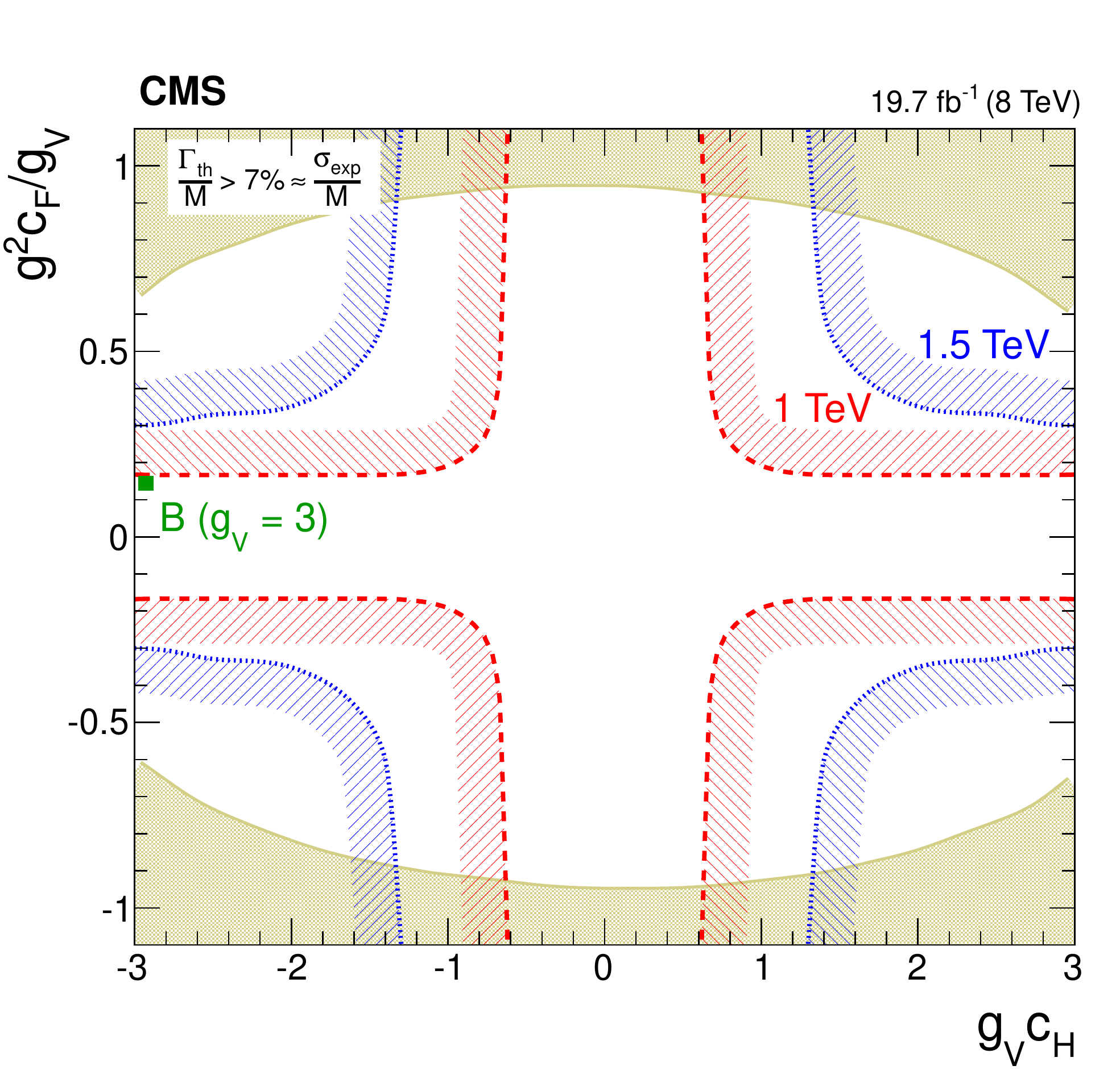}
\caption{Expected and observed upper limits on the quantity $\sigma$($Z^{\prime}$) B($Z^{\prime}$ $\rightarrow$ ZH) for the analysis channels combined (left). Exclusion regions in the plane of the HVT-model coupling constants ($g_{V}c_{H}, g_{2}c_{F}/g_{V}$) for two resonance masses, 1.0 and 1.5 TeV. }
\label{fig:fifth}
\end{figure}

Having observed no significant deviations in the observed number of events from the expected background, upper limits on the production cross section of a new resonance in the ZH final state has been set. In Fig~\ref{fig:fifth}, the parameters are chosen to be $g_{V}$ = 3 and $c_{F}$ = -$c_{H}$ = 1, corresponding to a strongly coupled sector and a scan of the coupling parameters and the corresponding regions of exclusion in the Heavy Vector Triplet (HVT) model are shown. The parameters are defined as $g_{V}c_{H}$ and $g_{2}c_{F}/g_{V}$, related to the coupling strength of the new resonance to the Higgs boson and to fermions. Regions of the plane excluded by this search are indicated by hatched areas. Ranges of the scan are limited by the assumption that the new resonance is narrow~\cite{Khachatryan:2015ywa}.

\subsection{Search for high-mass diphoton resonances at $\sqrt{s}$ = 8 TeV}

\begin{figure}[htbp]
\centering
\includegraphics[height=4.5cm]{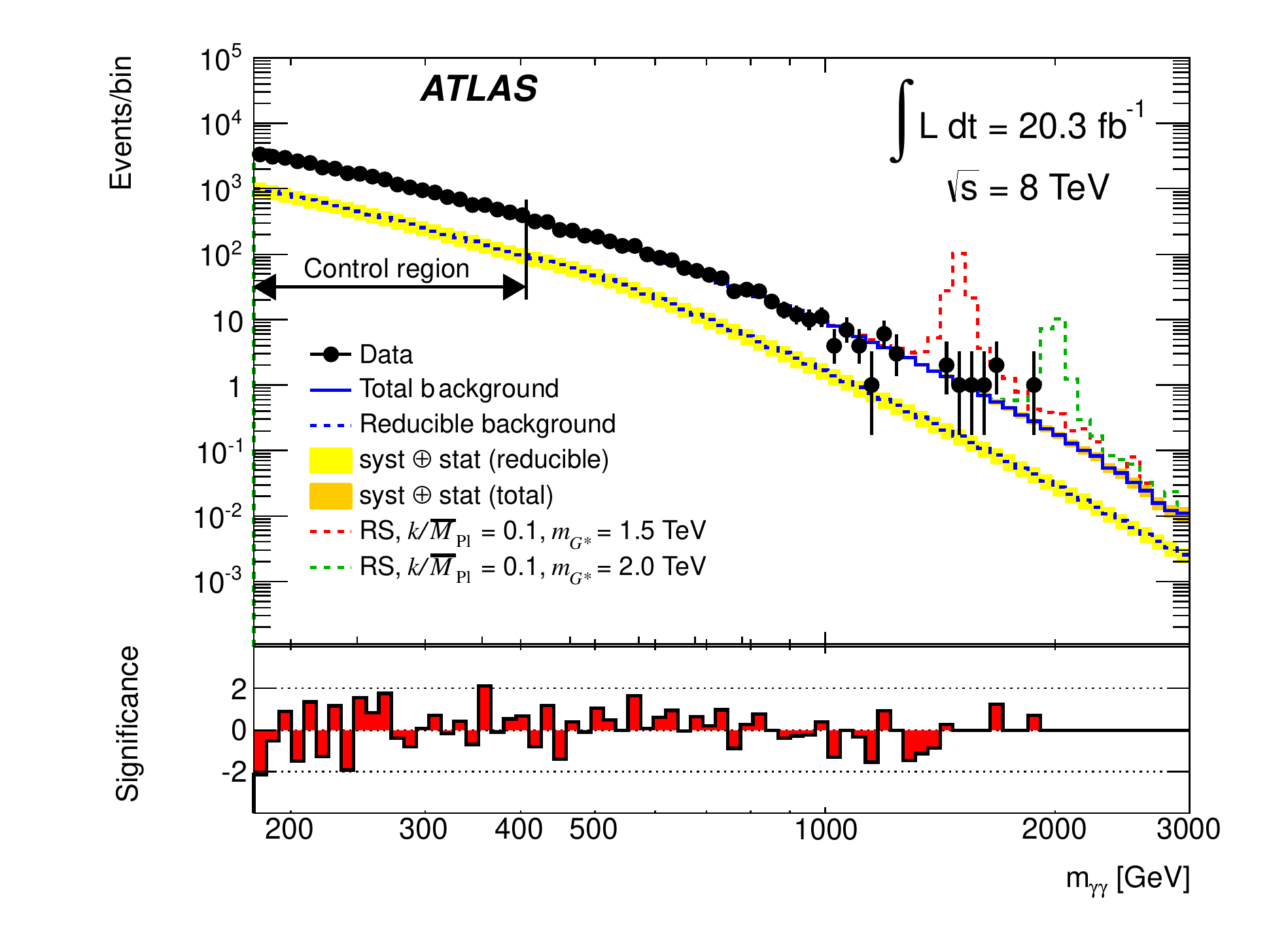}
\includegraphics[height=4.5cm]{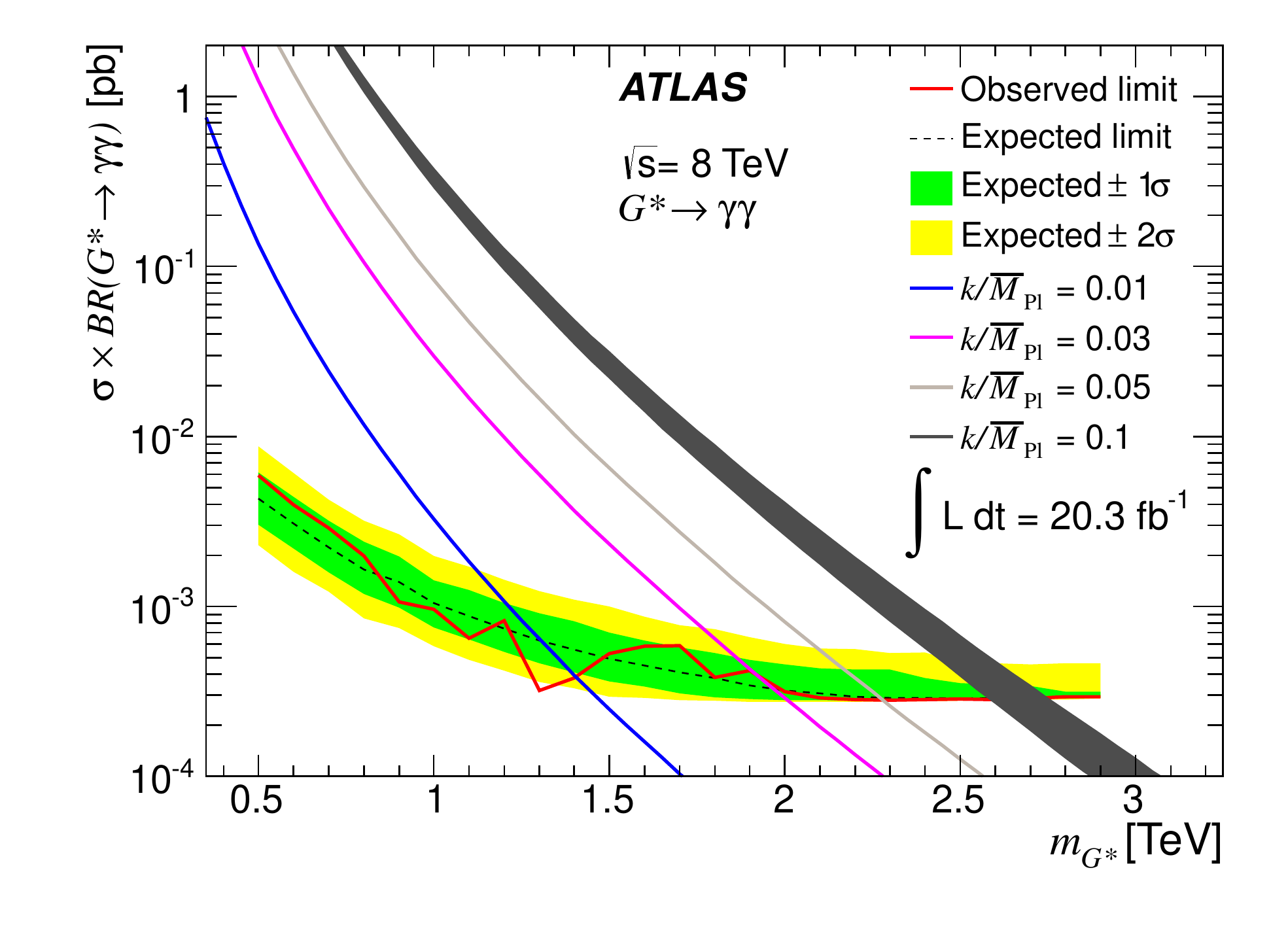}
\caption{Observed invariant mass distribution of the selected diphoton events (left).  95$\%$ CL expected and observed upper limits on $\sigma$ x BR($G^{*}$ $\rightarrow$ $\gamma\gamma$) as a function of the assumed value of the graviton mass (right).}
\label{fig:sixth}
\end{figure}

The diphoton search is another promising channel for the BSM physics. As is known, new high-mass states decaying to two photons are predicted in many extensions of the SM. In addition, the diphoton channel provides a clean experimental signature: excellent mass resolution and modest backgrounds. One popular example of BSM models predicting a new high-mass diphoton resonance is the Randall-Sundrum (RS) model~\cite{Randall:1999ee}. In this search, the observed diphoton mass spectrum together with the background expectation and the predicted signal for two examples of values of the RS model parameters has been shown in Fig.~\ref{fig:sixth}. No significant excess over the expected background is observed, and upper limits on the production cross section times branching fraction $\sigma$ x BR($G^{*}$ $\rightarrow$ $\gamma\gamma$) for narrow resonances are reported as a function of the resonance mass. 

\subsection{Search for the Production of Dark Matter in Association with Top Quark Pairs in the Single-lepton Final State at $\sqrt{s}$=8 TeV}

In this search, only one new Dirac fermion related to Dark Matter (DM) within the energy reach of the LHC has been considered. The fermion interacts with quarks via four-fermion contact interaction, which can be described by an effective filed theory (EFT). Assuming a DM particle with a mass of 100 GeV, the excluded interaction scale is 120 GeV for scalar interaction between top quarks and DM particles. Thus, search for the production of DM particles in association with a pair of top quarks, and consider only the scalar interaction, has been performed~\cite{Khachatryan:2015nua}.

\begin{figure}[htbp]
\centering
\includegraphics[height=4.7cm]{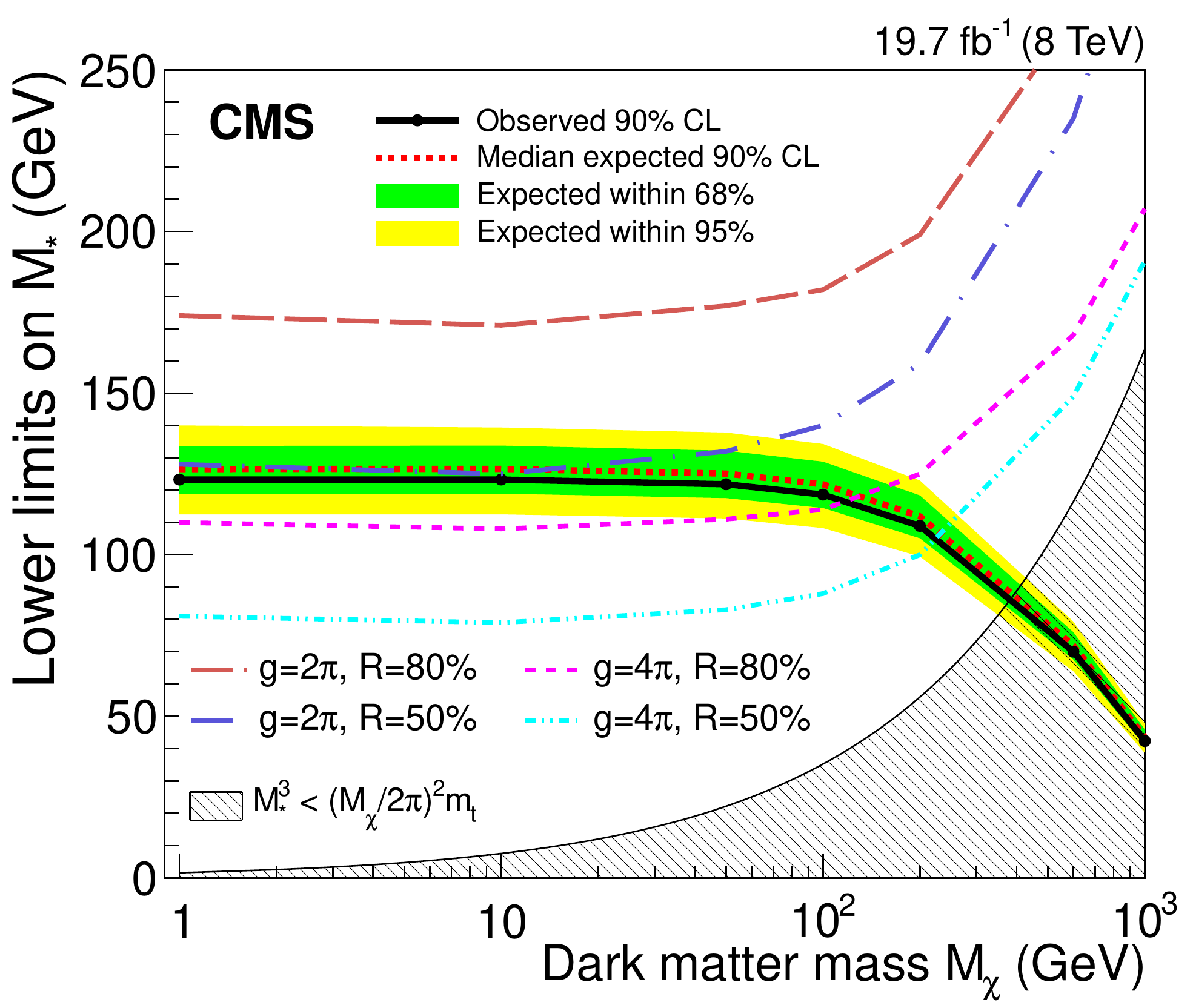}
\includegraphics[height=4.7cm]{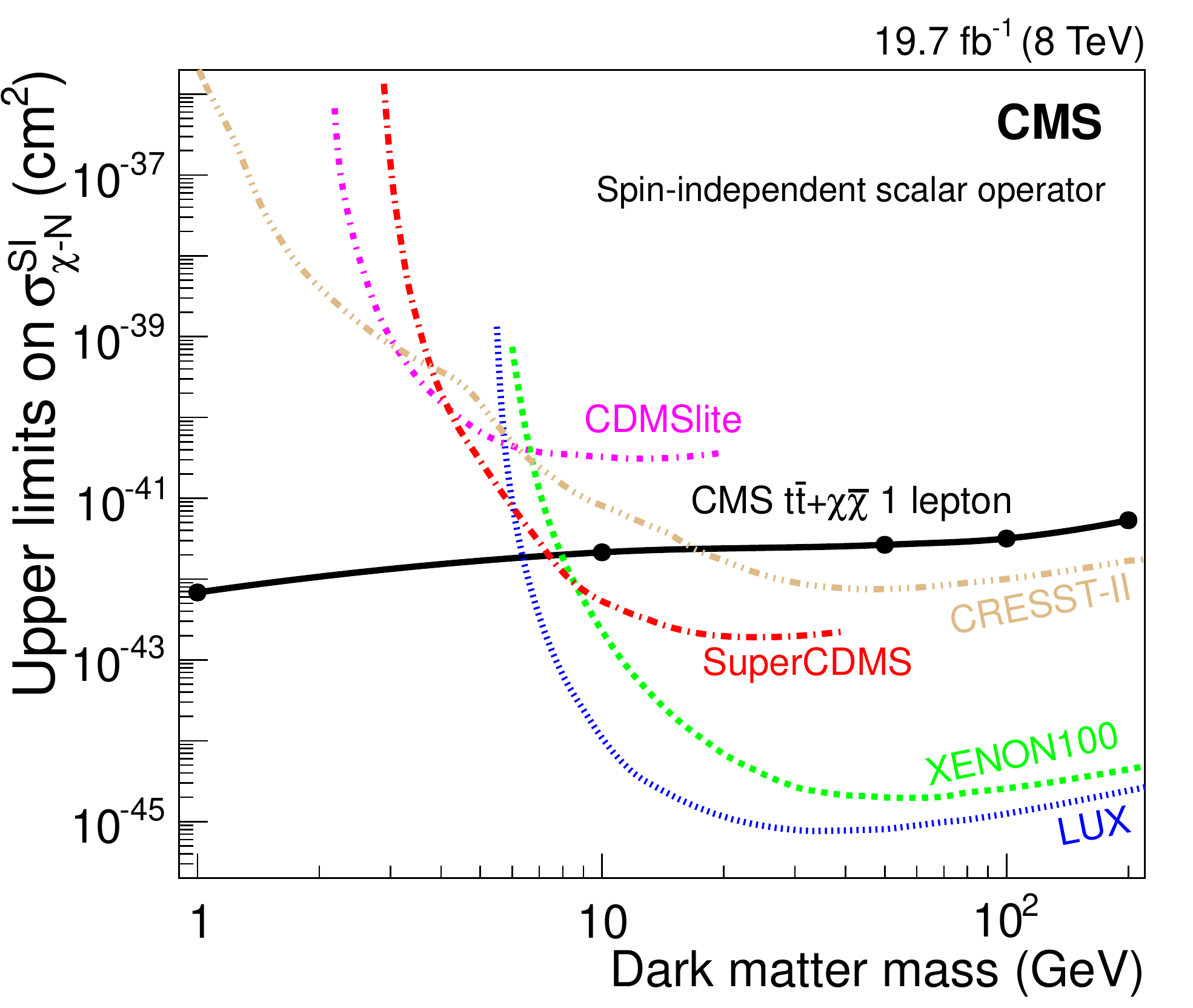}
\caption{90$\%$ CL observed exclusion limits in the plane of DM particle mass and interaction scale (left), and  DM-nucleon cross section as a function of the DM mass for the scalar operator (right). }
\label{fig:seventh}
\end{figure}

A lower bound of the validity of the EFT is indicated by the upper edge of the hatched area in Fig.~\ref{fig:seventh} (left). The four curves, corresponding to different g and R values, represent the lower bound on M* for which 50$\%$ and 80$\%$ of signal events have a pair of DM particles with an invariant mass less than g$\sqrt{M^{3}_{*}/m_{t}}$, where g = 4$\pi$ and g = 2$\pi$ respectively. On the other hand, the limits on the interaction scale $M_{*}$ can be translated to limits on the DM-nucleon scattering cross section~\cite{Khachatryan:2015nua}. Figure~\ref{fig:seventh}  (right) shows the observed 90$\%$ CL upper limits on the DM-nucleon cross section as a function of the DM mass for the scalar operator considered in this analysis.

\section{Summary}

Run-I at the LHC demonstrated the excellent performance and sensitivity over wide range of signatures but in fact just started to test various BSM physics. Constraints have been placed on models resulting in many SUSY and exotica scenarios. Many other analyses could also have sensitivity to models not yet considered, and could place bounds on new theories. 

Recently, Run-2 successfully started at a center-of-mass energy of 13 TeV - almost double that used for Run 1. This affects the current boundaries of the total integrated luminosity and the cross-sections of all SM and BSM physics processes including challenges of new era (TeV leptons, boosted objects and higher pile-up) at the CMS and ATLAS experiments. This will greatly improve the discovery potential of many beyond the Standard Model searches, giving a very optimistic outlook for the LHC Run 2.

\end{document}